\documentclass[prl,twocolumn,superscriptaddress]{revtex4}
\usepackage{times}
\usepackage{amsmath}
\usepackage{amsthm}
\usepackage{amssymb}
\usepackage{amsbsy}
\usepackage{graphicx}
\usepackage{pstricks}

\begin{document}

\newcommand{\smallrech}{ \; \pspicture[0.35](0.2,0.1)
\psset{linewidth=0.025,linestyle=solid}
\psline[](0.1,0)(0.1,0.1)
\pspolygon[](0,0)(0.2,0)(0.2,0.1)(0,0.1)
\endpspicture\;}

\newcommand{\smallrecv}{ \; \pspicture[0.35](0.1,0.2)
\psset{linewidth=0.025,linestyle=solid}
\psline[](0,0.1)(0.1,0.1)
\pspolygon[](0,0)(0,0.2)(0.1,0.2)(0.1,0.0)
\endpspicture\;}

\newcommand{\emptya}{  \; \pspicture[0.35](0.6,0.2)
\psdots[linecolor=black,dotsize=.10]  (0.15,0)(0.15,0.3)  (0.6,0.15)
\psset{linewidth=0.03,linestyle=dotted}
\psline[](0.3,0)(0.3,0.3)
\pspolygon[](0,0)(0.6,0)(0.6,0.3)(0,0.3)
\psset{linewidth=0.05,linestyle=solid}
\psline[linecolor=black](0,0)(0.3,0)
\psline[linecolor=black](0,0.3)(0.3,0.3)
\psline[linecolor=black](0.6,0)(0.6,0.3)
\endpspicture \;}

\newcommand{\emptyb}{ \; \pspicture[0.35](0.6,0.2)
\psdots[linecolor=black,dotsize=.10]  (0.45,0)(0.45,0.3)(0,0.15)
\psset{linewidth=0.03,linestyle=dotted}
\psline[](0.3,0)(0.3,0.3)
\pspolygon[](0,0)(0.6,0)(0.6,0.3)(0,0.3)
\psset{linewidth=0.05,linestyle=solid}
\psline[linecolor=black](0.3,0)(0.6,0)
\psline[linecolor=black](0.3,0.3)(0.6,0.3)
\psline[linecolor=black](0,0)(0,0.3)
\endpspicture \;}

\newcommand{\fulla}{ \; \pspicture[0.35](0.6,0.2)
\psdots[linecolor=black,dotsize=.10]   (0.15,0)(0.15,0.3)  (0.6,0.15)(0.3,0.15)
\psset{linewidth=0.03,linestyle=dotted}
\psline[](0.3,0)(0.3,0.3)
\pspolygon[](0,0)(0.6,0)(0.6,0.3)(0,0.3)
\psset{linewidth=0.05,linestyle=solid}
\psline[linecolor=black](0,0)(0.3,0)(0.3,0.3)(0,0.3)
\psline[linecolor=black](0.6,0)(0.6,0.3)
\endpspicture\;}

\newcommand{\fullb}{ \; \pspicture[0.35](0.6,0.2)
\psdots[linecolor=black,dotsize=.10]  (0.45,0)(0.45,0.3)  (0,0.15)(0.3,0.15)
\psset{linewidth=0.03,linestyle=dotted}
\psline[](0.3,0)(0.3,0.3)
\pspolygon[](0,0)(0.6,0)(0.6,0.3)(0,0.3)
\psset{linewidth=0.05,linestyle=solid}
\psline[linecolor=black](0.6,0)(0.3,0)(0.3,0.3)(0.6,0.3)
\psline[linecolor=black](0,0)(0,0.3)
\endpspicture \;}

% You should use BibTeX and apsrev.bst for references
\bibliographystyle{apsrev}
%\preprint{}

\title{Correlated fermions on a checkerboard lattice}
% Optional argument for running titles on pages
%\title[]{}



\author{Frank Pollmann}
\affiliation{Max-Planck-Institute for the Physics of Complex Systems, D-01187
Dresden, Germany}
\author{Joseph J. Betouras}
\affiliation{Instituut-Lorentz for Theoretical Physics, P.O. Box 9506,
NL-2300RA Leiden, The Netherlands}
\affiliation{School of Physics and Astronomy, %SUPA,
Scottish Universities Physics Alliance, 
University of St Andrews, North Haugh KY16 9SS, UK}
\author{Kirill Shtengel}
%\email[]{shtengel@physics.ucr.edu}
\affiliation{Department of Physics and Astronomy, University of California,
Riverside, CA 92521, USA}
\author{Peter Fulde}
\affiliation{Max-Planck-Institute for the Physics of Complex Systems, D-01187
Dresden, Germany} 

%\thanks{}
%\altaffiliation{}/19v
\date{\today}

\begin{abstract}
A model of strongly correlated spinless fermions on a
checkerboard lattice is mapped onto a quantum FPL model. We
identify a large number of fluctuationless states specific to the
fermionic case. We also show that for a class of fluctuating states,
the fermionic sign problem can be gauged away. This claim is supported
by numerical evaluation of the low-lying states. Furthermore,
we analyze excitations at the Rokhsar-Kivelson point of this
model using the relation to the height model and the single-mode
approximation.
\end{abstract}

%\maketitle must follow title, authors, abstract and \pacs
\maketitle
%%%%%%%%%%%%

%\paragraph{Introduction --}
The interplay between quantum and geometric frustration can result in many
unusual properties. From that point of view,
strongly correlated spinless fermions on a checkerboard
lattice present an interesting case. At certain fillings,
fractionally charged excitations have been predicted for the case of large
nearest-neighbor repulsion \cite{Fulde02}.
The subject is not purely academic since the checkerboard lattice is a
projection of a pyrochlore lattice onto a plane. There is experimental
evidence that electrons in pyrochlore lattices can be strongly
correlated  \cite{Kondo97}. We shall show that at half filling, this problem
can be mapped onto a
quantum fully packed loop (FPL) model -- an analog of the
quantum dimer model (QDM) -- and discuss a number of consequences. QDMs,
originally introduced in the context of quantum magnetism \cite{Rokhsar88},
became a focus of recent interest following the discovery of a
gapped liquid phase on a triangular lattice \cite{Moessner01a}. In
particular, it has been established that a gapped phase
with deconfined excitations exists in 2D for QDMs on non-bipartite lattices
\cite{Moessner01a,Nayak01a,Fendley02,Misguich02} while on bipartite lattices,
such as a square lattice, systems typically crystallize into a phase with a
broken translation/rotation symmetry. The liquid phase is ``shrunk'' into a
quantum critical point with gapless excitations
\cite{Rokhsar88,Moessner03:SMA,Fradkin04}. In both cases, an effective gauge
theory is a U(1) theory for such a critical point and a Z$_2$
theory for a deconfined phase \cite{Fradkin90,Moessner01b}. Several
related models, such as the quantum six/eight-vertex models
\cite{Ardonne04,Chakravarty02,Shannon04,Syljuasen06} have been
shown to conform to the same dichotomy: a model with orientable loops is
critical and is described by a U(1) gauge theory.

The aforementioned models share one important feature: the matrix elements
connecting various states of the low-energy Hilbert space are all
non-negative. The Hilbert space separates into different sectors so that all
states within a sector are connected by the quantum dynamics while states
belonging to different sectors are not. Then, by Perron-Frobenius theorem, the
ground state (GS) of  each sector is unique and nodeless
\cite{Moessner01a,Castelnovo05b}.
Much less is known about models with non-Frobenius dynamics which results in
some negative matrix elements. A striking difference between Frobenius vs.
non-Frobenius QDMs on the kagom\'{e} lattice was found in
\cite{Misguich02,Misguich03}: while the conventional QDM exhibits a gapped
Z$_2$ topological phase, its counterpart with different signs has an extensive
GS degeneracy. A different non-Frobenius model of
lattice fermions \cite{Fendley05b} has also been found to have an extensive
GS degeneracy.

\begin{figure}[thb]
\includegraphics[height=16mm]{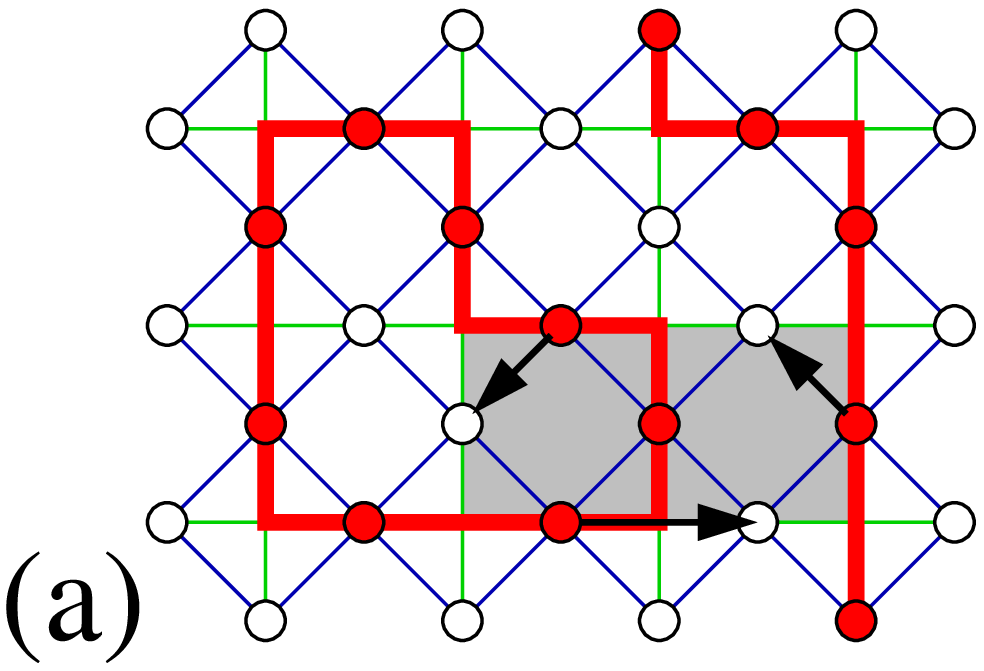}
$\;$
\includegraphics[height=16mm]{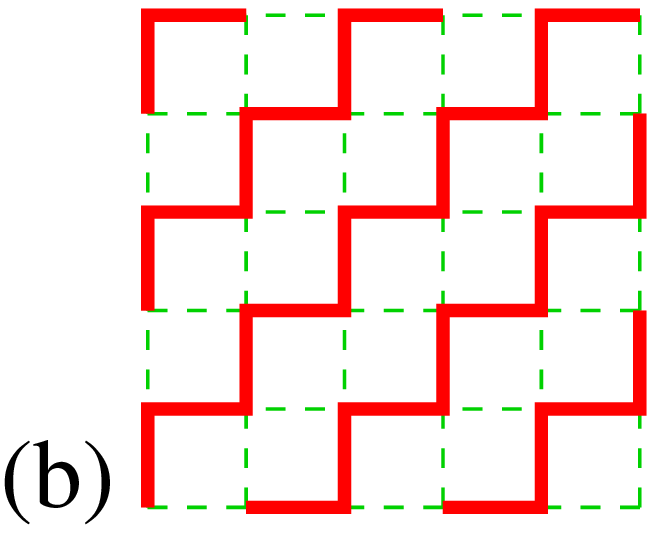}
$\;\;$
\includegraphics[height=16mm]{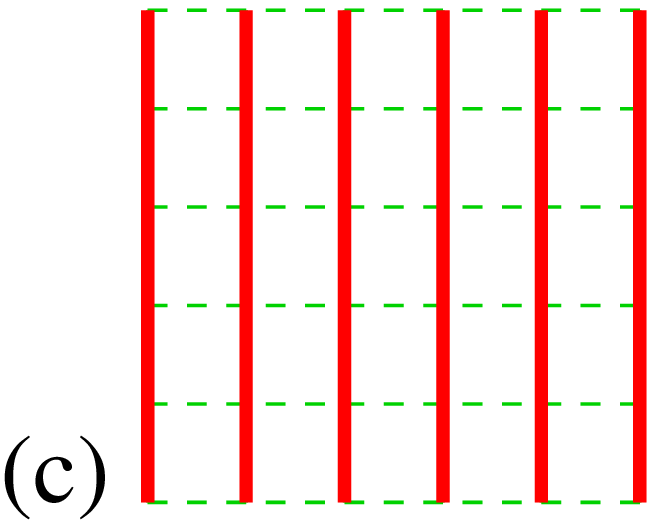}
\\
\vskip 3mm
\includegraphics[height=16mm]{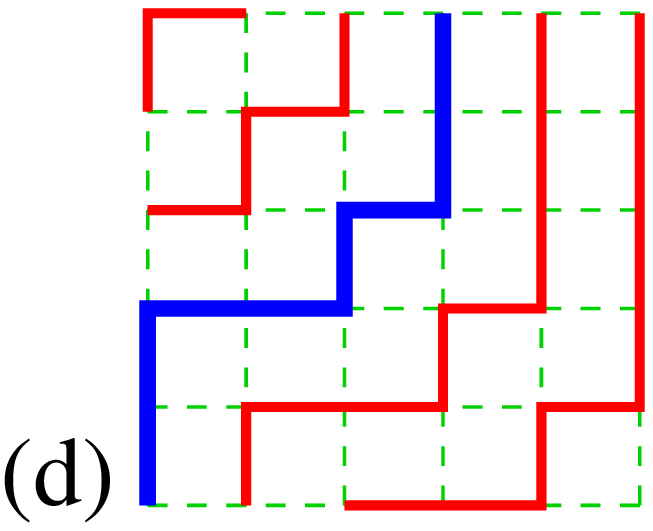}
$\quad$
\includegraphics[height=16mm]{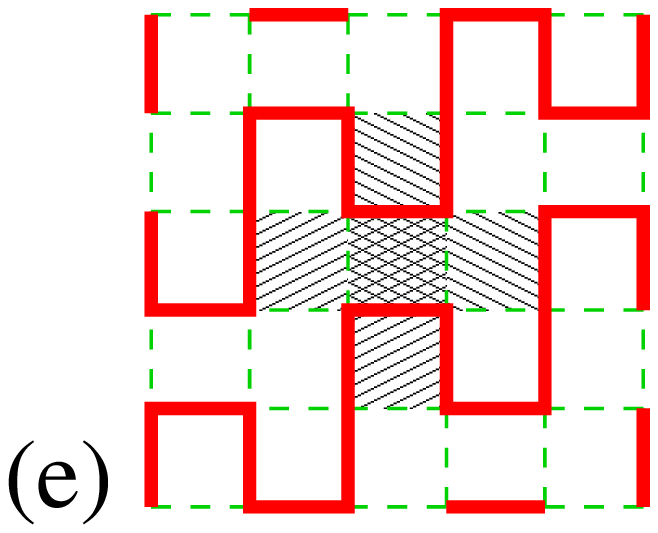}
$\quad$
\includegraphics[height=16mm]{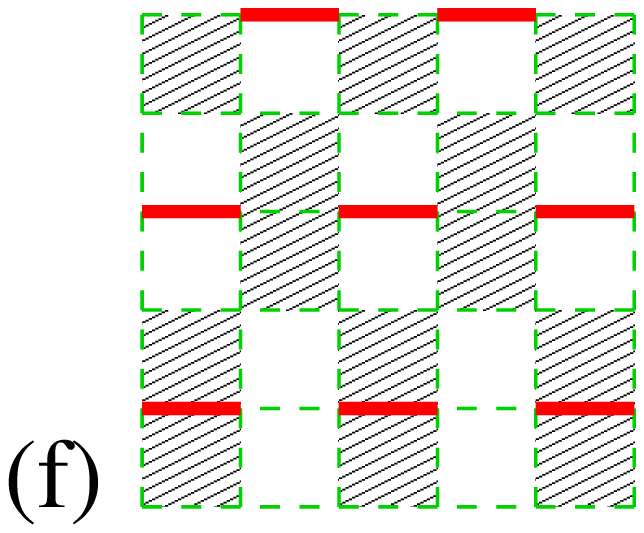}
\caption{(a) --
A configuration of fermions satisfying the ``two per tetrahedron''
rule and its representation in terms of FPLs.
A smallest ``flippable plaquette'' is shaded (see text for
details).
(b--d) -- Fragments of possible ``frozen'' states: (b)
is fluctuationless under \emph{any} local ring exchanges while (c) and (d) are
fluctuationless only for fermionic ring exchanges.
(e) -- The long-range ordered ``squiggle''
state maximizing the number of three-dimer ring exchanges (shaded). (f) --
Analog of a plaquette phase, with three dimers predominantly fluctuating
around shaded rectangles, shown dimers are not participating in these
exchanges.}
\label{fig:lattice-loops}
\end{figure}

The starting Hamiltonian for spinless fermions is:
%of the form
\begin{equation}
\label{eq:Hubbard1}
H=-t\sum_{\langle i\
j\rangle}\left(c_{i}^{\dag}c^{\vphantom{\dag}}_{j}+\text{H.c.}\right)+V\sum_{\langle i\
j\rangle}n_{i}n_{j},
\end{equation}
the summation is over neighboring sites of a checkerboard lattice. We assume
that $0<t\ll V$ -- the
limit of strong correlations. At half-filling which is studied
here, the potential term is minimized whenever there are two fermions on each
``planar tetrahedron'', i.e., a crisscrossed square in
Fig.~\ref{fig:lattice-loops}(a) (the tetrahedron rule, 
\cite{Anderson56}). Configurations satisfying this rule can be
represented by dimers on a square lattice connecting the centers of the planar
tetrahedra, with the particles sitting in the middle of the dimers
(Fig.~\ref{fig:lattice-loops}(a)) \cite{Pollmann06a}. Hence, the states
satisfying the tetrahedron rule are represented by non-intersecting FPL on the
square lattice. Different loop configurations are orthogonal to
each other as they correspond to different locations of fermions (in the
tight-binding limit, the overlap between the corresponding wavefunctions is neglected). 

We now turn our attention to the kinetic term in Hamiltonian (\ref{eq:Hubbard1}). It
creates configurations in which the tetrahedron rule is violated. But since we
are interested in the low-energy physics we may eliminate these
configurations and consider instead an effective ring-exchange Hamiltonian
\cite{Runge04} that acts \emph{within} the FPL Hilbert subspace. To lowest
non-vanishing order in $t/V$, such effective Hamiltonian becomes:
\begin{multline}
H_{\text{eff}} = g \sum_{\{\smallrech,\smallrecv\}}
\Big( \big|\emptya \big\rangle \big\langle\emptyb \big|+\text{H.c.}\Big)
\\
- g \sum_{\{\smallrech,\smallrecv\}}
\Big( \big|\fulla \big\rangle \big\langle\fullb \big|+\text{H.c.}\Big)
\\
\equiv g\sum_p \Big(\big| A \big\rangle\big\langle \overline{A} \big| + \big| \overline{A} \big\rangle\big\langle {A} \big|
- \big| B \big\rangle\big\langle \overline{B} \big| - \big| \overline{B} \rangle\langle {B} \big|\Big)
\label{eq:ringexone}
\end{multline}
where the sums are performed over all polygons of perimeter six and $g = 12
t^3/V^2$. Particles are located in the middle of the dimers. The different
signs result from the number of permuted fermions which is either even or
odd. 
%%%%%%%%%%%%%%%%%%%%%%%%%%%%%%%%%%%%%%%%%%%%%%%%%%%%%%%%%%%%%%%%%%%%%%%%%%%
%%%%%%%%%%%%%%%%%%%%%%%%%%%%%%%%%%%%%%%%%%%%%%%%%%%%%%%%%%%%%%%%%%%%%%%%%%%

\paragraph{Symmetries and Conserved Quantities --}

The first important point is that this model is a variation of
the quantum six-vertex (6V) model \cite{Chakravarty02,Ardonne04,Syljuasen06}
\footnote{The
mapping between FPL and 6V is done by subdividing the square lattice into even
and odd sublattices and orienting occupied bonds from odd to even sites, with
the opposite orientation for vacant bonds.}.
However, there are two important differences resulting from the fermionic
statistics. Firstly,  only ring exchanges involving \emph{odd numbers} of
particles are allowed; the contributions from clock- and counterclockwise ring
moves of an even number of fermions cancel. In particular, the smallest
resonating plaquette has perimeter six,
rather than four, as captured by Eq.~(\ref{eq:ringexone}). Secondly, the
sign of the ring-hopping term depends on the
occupancy of the middle bond. Both differences have important
implications: the first one -- for the ergodicity of the quantum dynamics,
while the second one leads to a sign problem in the
quantum Monte Carlo dynamics.

Let us now discuss symmetries and conservation laws pertaining to our model.
Firstly, since its Hilbert space is equivalent to that of a 6V
model, it has a height representation \cite{van_Beijeren77}. The quantum
dynamics can also be described in the height language
\cite{Henley97,Henley04}. The immediate consequence is that \emph{no}
such local dynamics can change the global slopes
$\kappa_x(y)$ and $\kappa_y(x)$ in $x$ and $y$ directions, which therefore
form a set of conserved numbers. Additionally, there is an accidental symmetry
in our model, specific to the
quantum dynamics (\ref{eq:ringexone}): it conserves the total numbers of both
vertical and horizontal dimers (moreover, it conserves the number of
horizontal dimers in even and odd rows separately, and similarly for the
vertical dimers). Hence, all states can be classified with respect to these
quantum numbers which are \emph{different} from the aforementioned conserved
slopes. Notice that these additional conservation laws will not survive if
higher-order ring-exchange terms are added to the Hamiltonian
(\ref{eq:ringexone}).

% Finally, we note that the overall sign of $H_{\text{eff}}$ (i.e.,
% the sign of $g$) in Eq.~(\ref{eq:ringexone}) is a matter of convention (as
% has been first pointed out in \cite{Rokhsar88} for the related QDM). 
% Introducing a factor of $i$ for each vertical dimer in an odd row of squares
% and for each horizontal dimer in an odd column changes this sign.

\begin{figure}[bht]
\includegraphics[width=60mm]{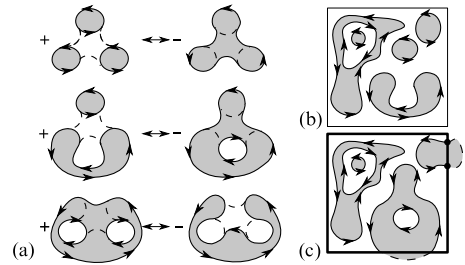}
\caption{(a) Three different actions of the effective Hamiltonian on the
topology of loop configurations. (b, c) Representation of configuration as
fully-packed directed loops.}
\label{fig:loops}
\end{figure}

\paragraph{Sign Problem --}
The sign problem manifested in the opposite signs of the $A\leftrightarrow
\overline{A}$ and $B\leftrightarrow \overline{B}$ terms in
Eqs.~(\ref{eq:ringexone}) can be avoided in certain (but not all) cases.
Notice that $B\leftrightarrow \overline{B}$ processes do not change the loop
topology (or their number), while the $A\leftrightarrow \overline{A}$ flips
always do, with the possibilities presented schematically in
Fig.~\ref{fig:loops}(a). Let us consider the coarse-grained surface with an
even number of fermions remaining fixed at the boundary (``fixed'' boundary
conditions (BC)). Then fermion configurations are represented by closed loops
as well as arcs terminating at the boundary. With no fermions at the
 boundary, there are closed loops only (Fig.~\ref{fig:loops}(b)), which we 
orient as follows: (i) color the areas separated by the loops white
and grey, with white being the outmost color; (ii) orient all loops so that
the white regions are always to the right. In the presence of arcs
(Fig.~\ref{fig:loops}(c)) we can choose how to close them on the
outside without intersections (the outside region has no dynamics). Letting
white be the color at infinity, we orient the loops as described above.
We now notice that the relative signs resulting from the $A\leftrightarrow
\overline{A}$ flips are consistent with multiplying each loop
configuration by $i^l (-i)^r$, where $r$
($l$) is the number of the (counter-) clockwise winding loop
 Hence, by simultaneously changing the sign of the $A\leftrightarrow
\overline{A}$ flips and transforming the loop states $|\mathcal{L} \rangle \to
i^{l(\mathcal{L})} (-i)^{r(\mathcal{L})} |\mathcal{L} \rangle$, we cure the
sign problem thus making the system effectively bosonic. (In fact, this fix
can be related to a more general class of gaugeable signs
discussed in \cite{Castelnovo05b}.)
%%%%%%%%%%%%%%%%%%%%%%%%%%%%%%%%%%%%%%%%%%%%%%%%%%%%%%%%%%%%%%%%%%%%%%%%%%%
%%%%%%%%%%%%%%%%%%%%%%%%%%%%%%%%%%%%%%%%%%%%%%%%%%%%%%%%%%%%%%%%%%%%%%%%%%%

%%%%%%%%%%%%%%%%%%%%%%%%%%%%%%%%%%%%%%%%%%%%%%%%%%%%%%%%%%%%%%%%%%%%%%%%%%%
\begin{figure}[thb]
\includegraphics[width=68mm]{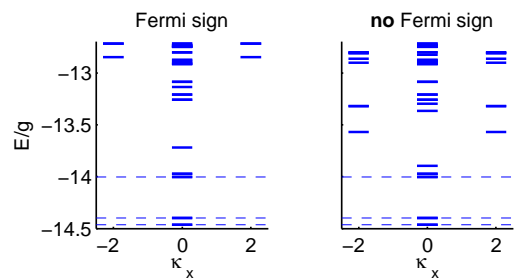}
\caption{GS energy and energies of lowest excited states of the
effective Hamiltonian in sectors with different slopes ($\kappa_x$, $0$). Left
panel: Data from an exact diagonalization of $H_{\text{eff}}$ on a 72 site
cluster where the Fermi sign is taken into account. Right panel: Same data
from a calculation with excluded Fermi signs.}
\label{fig:parabola}
\end{figure}
%%%%%%%%%%%%%%%%%%%%%%%%%%%%%%%%%%%%%%%%%%%%%%%%%%%%%%%%%%%%%%%%%%%%%%%%%%%
Note that this construction need not work for the periodic BC: firstly, only
even-winding sectors on a torus allow two-coloring, and secondly, even for
even windings, it is possible to dynamically reverse the coloring while
returning to the same loop configuration, resulting in
non-trivial state counting.
It appears though that for the periodic BC on \emph{even} tori (preserving
the bipartiteness of the lattice), the lowest-energy states belong to the
sector where such transformation works, as confirmed by the exact
diagonalization results (Fig.~\ref{fig:parabola}). The
presented non-local loop-orienting construction is restricted to the
ring-exchange processes of length six.

\paragraph{Rokhsar-Kivelson (RK) Point --}

It has been shown that at half-filling the Hamiltonian (\ref{eq:ringexone})
has an ordered GS and fractionally charged, weakly confined excitations
\cite{Pollmann06b}. 
In the spirit of \cite{Rokhsar88}, we extend our model's parameter space in
order to study its phase diagram and consider the Hamiltonian
\begin{multline}
 \label{eq:ringextwo}
H_{\text{eff}}^{\prime} = \sum_p \left[g\left(\left| A \rangle
\langle \overline{A} \right| + \left| \overline{A} \rangle
\langle {A} \right|
- \left| B \rangle
\langle \overline{B} \right| - \left| \overline{B} \rangle
\langle {B} \right|\right) \right.
\\
\left.
+ \mu \left( \left| A \rangle
\langle {A} \right| + \left| \overline{A} \rangle
\langle \overline{A} \right|
+ \left| {B} \rangle
\langle {B} \right| + \left| \overline{B} \rangle
\langle \overline{B} \right|\right)\right]
\end{multline}
where we use the short-hand notations introduced in Eq.~(\ref{eq:ringexone}).
At the RK point $g=\mu$, this Hamiltonian can be written as a sum of
projectors: 
\begin{multline}
 \label{eq:RK}
H_{\text{eff}}^{\prime}|_{g=\mu} = g\sum_p \left[\left( \left| A \rangle
+ \left| \overline{A} \rangle \right)
\left(
\langle  A \right| + \langle  \overline{A} \right| \right)
\right.
\\
+ \left.
\left( \left| B \rangle - \left| \overline{B}
\rangle \right)
\left(
\langle  B \right| - \langle  \overline{B} \right| \right)
\right].
\end{multline}
A (non-unique) GS of Hamiltonian (\ref{eq:RK}) is a state that is annihilated
simultaneously by all projectors. There are two kinds of such states:
\emph{frozen} states and \emph{liquid} states. 
It turns out that the fermionic nature has strong consequences for both kinds.
All bosonic frozen states identified in \cite{Shannon04} remain frozen since
they are unaffected by \emph{any} local ring exchange. In the height language,
these states maximize the slope in either $x$ or $y$ direction, making any
local dynamics impossible. Fixing this slope in an $L\times L$ system, one
still has $2^L$ possible arrangements of steps in the other direction which
results in $4\times 2^L$ such states \footnote{This is unlike the QDM case
with only 4 such frozen, or \emph{staggered} states: maximizing the slope in
one direction there automatically fixes it in the other.}. However, there are
also new additional GSs (see Fig.~\ref{fig:lattice-loops}(c,d)) which remain
unaffected by any \emph{fermionic} dynamics but would not be so in the bosonic
case. This is due to the fact that only odd-fermion ring exchanges are allowed
here. For the types of states shown in Fig.~\ref{fig:lattice-loops}(c,d), this
is easy to see: any ring exchange is an alternating sequence of occupied and
vacant bonds. The subsequence of vacant bonds maps onto a 1D walk between the
lines and must contain an \emph{even} number of steps in order to close. The
states of this type are obtained by translating a single line like in
Fig.~\ref{fig:lattice-loops}(d), e.g., a line starting at the left bottom 
corner $(0,0)$ and reaching the top boundary at $(x,L)$ while going only up
and to
the right (we assume free BC) can do it in ${L+x}\choose{x}$
distinct ways leading to $4\times \sum_{x=0}^{L} {{L+x}\choose{x}} \to
4^{L+1}$ (for large $L$) such frozen fermionic states. There is also a number
of other frozen fermionic configurations that we could pinpoint (but not
systematically count). Hence it remains an open issue whether our RK point is
characterized by an extensive GS entropy and whether such entropy observed in
\cite{Misguich03,Fendley05b} can be explained in this manner. However, it is
very suggestive that the explanation lies in a more restrictive nature of the
fermionic dynamics. We stress that these ``new'' frozen GSs are not related to
any height constraints -- in fact, the states of type shown in
(Fig.~\ref{fig:lattice-loops}(d)) include both the maximally tilted state
(Fig.~\ref{fig:lattice-loops}(b)) and the maximally flat state
(Fig.~\ref{fig:lattice-loops}(c)). ``Freezing'' of the latter state is
especially interesting since it is supposed to dominate the ordered phase of
the bosonic model for $\mu/g < -0.374$ \cite{Shannon04} by maximizing the
number of the flippable plaquettes  \footnote{This is the conventional
anti-ferroelectric phase of the six-vertex model referred to as the N\'{e}el
phase in \cite{Shannon04} and as DDW phase in
\cite{Chakravarty02,Syljuasen06}}. This observation puts into doubt the
existence of such a phase in the fermionic model.

In addition to the frozen states, there are also liquid-like GSs,  as was the
case in
the original RK construction. These states can be found exactly for the
dynamics given by Eqs.~(\ref{eq:ringexone}--\ref{eq:RK}) with the help of
 the aforementioned sign-fixing transformation. (While this transformation
changes the sign of the $A\leftrightarrow \overline{A}$ flips, it does not
affect the diagonal terms in Eqs.~(\ref{eq:ringextwo}).) Hence, for ``fixed''
BC, the state
 $|0\rangle \propto
\sum_{\{\mathcal{L}\}}
i^{l(\mathcal{L})} (-i)^{r(\mathcal{L})} |\mathcal{L} \rangle$ is
automatically annihilated by all projector terms in $H_{\text{eff}^{\prime}}$
and therefore is a GS. Here the sum is taken over all loop configurations
$\mathcal{L}$ that are connected to each other by the quantum dynamics. Note
that there are many disconnected sectors.

\paragraph{Excitations at the RK point --}
The quantum dynamics of the Hamiltonian $H_{\text{eff}}$ and similarly
$H_{\text{eff}}^{\prime}$ can
be described in terms of a height model \cite{Henley97,Henley04}. The
associated conserved quantities $\kappa_x$ and $\kappa_y$, as well as their
insensitivity to a constant shift of the height field imply gapless
hydrodynamic modes with $\omega(k)\sim k^2$ as $k\rightarrow 0$. The liquid
state of the FPL model at the RK point corresponds to the rough phase of the
height model, in which case the modes can be identified as the
so-called resonons at wavevector $(\pi, \pi)$ \cite{Rokhsar88} and the
equivalent of the pi0ns, here at $\mathbf  Q = (0,0)$ \cite{Moessner03b,
Moessner03a,Moessner04a}.

For constructing gapless excitations in our model, we adopt the single-mode
approximation which has been successfully used for the QDM on the square
lattice \cite{Rokhsar88,Moessner03b}. Let $|0\rangle$ be the aforementioned
``liquid'' GS at the RK point. The operator
$d_{\hat{\tau}}^{\pm}(\mathbf r)$  creates (annihilates) a dimer at position
$\mathbf r$ in direction $\hat{\tau}$. The density operator
$n_{\hat{\tau}}(\mathbf r)=d_{\hat{\tau}}^+(\mathbf r)d_{\hat{\tau}}^-(\mathbf
r)$ has the Fourier transform $n_{\hat{\tau}}(\mathbf q)=\sum_{\mathbf r}
n_{\hat{\tau}}(\mathbf r) \exp(i\mathbf q\cdot \mathbf r)$. We use the
operators $n_{\hat{\tau}}(\mathbf q)$  to construct the states $|\mathbf q,
\hat{\tau}\rangle=n_{\hat{\tau}}(\mathbf q)|0\rangle$ which are for $\mathbf
q\ne 0$ orthogonal to $|0\rangle$. The excitation energies are bounded by
$E(\mathbf q, \hat{\tau}) \leq f(\mathbf{q})/S_{\tau \tau}(\mathbf q)$, where
$f(\mathbf q)=\langle 0|[n_{\hat{\tau}}(-\mathbf q),[H_{\text{eff}}',
n_{\hat{\tau}}(\mathbf q)]]|0\rangle$ is the  the oscillator strength and
$S_{\tau\tau}(\mathbf q)=\langle 0|n_{\hat{\tau}}(-\mathbf q)
n_{\hat{\tau}}(\mathbf q)|0 \rangle$ is the structure factor. In order to
calculate $f(\mathbf q)$, we observe that the density operators commute with
the potential energy term, hence only the kinetic energy term contributes. By
using  the commutation relation $[d_{\hat{\tau}}^{\pm},n_{\hat{\tau}}]= \mp
d_{\hat{\tau}}^{\pm}$ repeatedly, we compute $f(\mathbf q)$ due to the
resonating terms for $\hat{\tau}=\hat x$.
The upshot is that to leading order $f(\mathbf k)\sim ({\mathbf k} \times
\hat{\tau})^2$ (similar type of expression as the one for the RK model
\cite{Rokhsar88}) with $\mathbf{k}=\mathbf{q}-\mathbf{Q}$ at $\mathbf{Q} =
(0,0), (\pi,\pi), (0,\pi)$. The second ingredient, the structure factor
$S_{xx}(\mathbf q)$ is the Fourier transform of the dimer density correlation
function. We use the expression given in Ref.~\cite{Moessner04a} for it.
Interestingly, $S_{xx}(\mathbf q)\ne 0$ except along the direction $\mathbf
q=(q_x,\pi)$ where it vanishes with the exception of $(\pi,\pi)$. At $(0,
\pi)$ both $f(\mathbf q)$ and $S_{xx}(\mathbf q)$ are zero, but their ratio
remains finite. In addition $S_{xx}(\mathbf q)$ shows no singularities. The
FPL model differs from the hardcore dimer model (describing our model at
quarter-filling) for which $S_{xx}(\mathbf q)$ diverges logarithmically at
$\mathbf Q=(\pi,0)$ \cite{Moessner03b}. The difference is due to a slower
algebraic decrease with distance of the correlation function for hardcore
dimer covering. We have also verified the above result for our structure
factor $S_{xx}(\mathbf q)$ by means of Monte Carlo simulations.

%\paragraph{Conclusions --}
We conclude with several observations and open questions addressing the
possible phase diagram of the model defined by Eq.~(\ref{eq:ringextwo}). The
bosonic version studied in \cite{Shannon04} has three phases depending on the
parameter $\mu/g$: the two ``flat'' phases, i.e., the N\'{e}el phase
\cite{Shannon04} and the plaquette phase (also found in
\cite{Moessner04a,Syljuasen06}) as well as a maximally tilted frozen phase,
with the RK point perched between the latter two phases. We have already
argued that some of the flat states are actually frozen for fermions and hence
cannot be used to define the fluctuation-stabilized phases for  $\mu/g<1$. The
nature, and number of such phases are an open question, but let us
attempt an analysis. Firstly, the maximally-flat (N\'{e}el, DDW) phase shown
in Fig.~\ref{fig:lattice-loops}(c) appears to be replaced by the somewhat
analogous ``squiggle'' phase
(Fig.~\ref{fig:lattice-loops}(e))\cite{Pollmann06b}. In the bosonic case, the
region of  $-0.374<\mu/g < 1$ corresponded to the plaquette phase, while its
direct fermionic analog does not appear to be present anywhere in the
fermionic phase diagram. (Notice, however, that unlike the N\'{e}el and the
plaquette phases, both states shown in Fig. ~\ref{fig:lattice-loops}(e,f)
break \emph{both} translation and rotation symmetries simultaneously.) We rule
this phase out based on the fact that the squiggle phase breaks the symmetry
between the numbers of the vertical and the horizontal dimers (which are
separately conserved by the quantum dynamics), while the ``fermionic
plaquette'' phase does not. The results of the exact diagonalization on the
small samples indicate that this symmetry remains broken all the way up to
$\mu/g = 1$ -- the RK point, with the ratio of 2/3 being consistent with the
``squiggle'' phase. This observation is also strongly disfavoring a critical 
liquid phase to the left of the RK point, hence the
fermionic RK point in this model is likely to be an isolated quantum critical
point just as it is for the bosonic model. The gapless modes identified above
are in agreement with such a scenario. The details of the phase
diagram and its sensitivity to the higher order ring exchanges is a
subject for future work.

%In summary, we have analyzed a model of strongly correlated spinless fermions
%on a half-filled checkerboard by mapping it onto a quantum fully-packed loop
%model on the square lattice with the kinetic energy given by ring exchanges.
%We showed that the energy of the ground-state and the lowest excited sates
%remains unchanged when the ring hopping term in $H_{\text{eff}}$ acts on
%bosons or fermions. Furthermore, we introduced a Hamiltonian
%$H^{\prime}_{\text{eff}}$ which differs from  $H_{\text{eff}}$ by an extra
%term allowing for a RK point at which the ground-state can be found exactly.
%From the above we conclude that at that point one hydrodynamic mode is at
%$\mathbf Q = (\pi,\pi)$ (resonon \cite{Rokhsar88}) while another one is at
%$\mathbf Q = (0,0)$. The latter corresponds to the pi0n for hardcore dimer
%covering \cite{Moessner03:SMA}. Given the presence of these soft modes the RK
%point must be critical.

%%%%%%%%%%%%%%%%%%%%%%%%%%%%%%%%%%%%%%%%%%%%%%%%%%%%%%%%%%%%%%%%%%%%%%%%%%%
%\begin{figure}[hbt]
%\includegraphics[width=60mm]{structure_factor.eps}
%\caption{Structure factor $S_{xx}$ for the loop covering on the square
%lattice.}
%\label{structurefactor}
%\end{figure}
%%%%%%%%%%%%%%%%%%%%%%%%%%%%%%%%%%%%%%%%%%%%%%%%%%%%%%%%%%%%%%%%%%%%%%%%%%%

\begin{acknowledgments}
The authors would like to thank R.~Moessner and E.~Runge for many
illuminating discussions.
F.~P.\ has benefited from the hospitality of IDRIS, Orsay during a stay made
possible by a HPC Europe grant (RII3-CT-2003-506079). K.~S.\ is grateful for
the hospitality of the MPIPKS, Dresden and the Instituut-Lorentz, Leiden.
While this paper was in preparation, we learned about a similar effort
undertaken by C.~Penc and N.~Shannon. 
\end{acknowledgments}
\bibliographystyle{apsrev}
\bibliography{./corr}

\end{document}